\documentclass[a4paper, 11pt]{article}
\usepackage[colorlinks=true, pdfstartview=FitV, linkcolor=blue, citecolor=red, urlcolor=magenta]{hyperref}
\usepackage{graphicx}
\usepackage{latexsym}
\usepackage{amsmath}
\usepackage{appendix}
\usepackage{float}
\usepackage{dsfont}
\usepackage{amsthm,esint}
\usepackage[makeroom]{cancel}
\usepackage{amsfonts}
\usepackage{bbm}
\usepackage{amssymb}
\usepackage{slashed}
\usepackage{verbatim}
\usepackage{braket}
\usepackage{color}
\usepackage{etoolbox}

\newcommand{\be}{\begin{equation}}
\newcommand{\ee}{\end{equation}}

\newcommand{\pa}{\partial}

\newcommand{\nn}{\nonumber}

\def\mc{\mathcal}

\newcommand{\beq}{\begin{eqnarray}}
\newcommand{\eeq}{\end{eqnarray}}

\usepackage{color}
\usepackage{etoolbox}
\patchcmd{\footnotemark}{\stepcounter{footnote}}{\refstepcounter{footnote}}{}{}
\definecolor{verde}{cmyk}{.83,.21,1,.08}

\newcommand{\R}{\mathbb R}

\def\beqa{\begin{eqnarray}}
\def\eeqa{\end{eqnarray}}
\def\a{\alpha}

\def\d{\delta}

\def\m{\mu}
\def\n{\nu}

\def\l{\lambda}
\def\o{\omega}
\def\p{\pi}
\def\r{\rho}
\def\s{\sigma}
\def\t{\tau}

\def\Th{\Theta}

\def\O{\Omega}

\def\bc{\bar{c}}

\newcommand{\eqn}[1]{(\ref{#1})}
\def\nn{\nonumber}

\def\appendix#1{\addtocounter{section}{1}\setcounter{equation}{0}
\renewcommand{\thesection}{\Alph{section}}
\section*{
\thesection\protect\indent \parbox[t]{11.715cm} {#1}}
\addcontentsline{toc}{section}{Appendix\thesection\ \ \ #1} }

\def\nn{\nonumber}

\def\bea{\begin{eqnarray}}
\def\eea{\end{eqnarray}}

\newcommand{\del}{\partial}

\newcommand{\dd}{{\mathrm d}}

\begin{document}

\begin{titlepage}


\begin{center}

\baselineskip=24pt

{\Large\bf Gribov horizon in Noncommutative QED}

\baselineskip=14pt

\vspace{1cm}

Ozório Holanda$^1$, Marcelo S. Guimaraes$^2$,  Luigi Rosa$^{3,5}$ and  Patrizia Vitale$^{4,5}$ 
\\ [6mm]
$^1${\it Centro de Matemática, Computação e Cognição - Universidade Federal do ABC, Santo André, SP, Brazil}\\
[4mm]
$^2$ {\it Departamento de F\'{\i }sica Te\'{o}rica, Instituto de F\'{\i }sica, UERJ - Universidade do Estado do Rio de Janeiro.\\ Rua S\~{a}o Francisco Xavier 524, 20550-013 Maracan\~{a}, Rio de Janeiro, Brasil.}\\
[4mm]
$^3${\it Dipartimento di Matematica e Applicazioni ``R.~Caccioppoli'', Universit\`{a} di Napoli
{\sl Federico II}}\\
{\it Monte S.~Angelo, Via Cintia, 80126 Napoli, Italy}
\\[4mm]
$^4${\it Dipartimento di Fisica ``E.~Pancini', Universit\`{a} di Napoli
{\sl Federico II}}\\
{\it Monte S.~Angelo, Via Cintia, 80126 Napoli, Italy}
\\[4mm]
$^5${\it INFN, Sezione di Napoli}\\
{\it Monte S.~Angelo, Via Cintia, 80126 Napoli, Italy}
\vskip 1truecm
{\small\tt 
netoholanda91@gmail.com, msguimaraes@uerj.br,  luigi.rosa@unina.it, patrizia.vitale@unina.it}

\end{center}

\begin{abstract}
It is known that  Noncommutative QED (NCQED) exhibits Gribov ambiguities in the Landau gauge. These ambiguities are related to zero modes of the Faddeev-Popov operator and arise in the ghost propagator when it has a pole. In this work, we establish a positive Faddeev-Popov operator for NCQED and the condition for the ghost propagator  not to have  poles, the so-called Gribov no-pole condition. This condition is implemented in the path integral and allows for the calculation of the photon propagator in  momentum space, which is dependent on the squared non-commutativity parameter. In the commutative limit, the standard QED is recovered.
\end{abstract}

\end{titlepage}


\section{Introduction}
The occurrence of so-called  Gribov ambiguities for Yang-Mills theories was first discussed in \cite{Gribov1978}, where Gribov showed that for non-Abelian gauge theories on flat topologically trivial space-times,   gauge fixing is problematic. Namely, it is not possible to choose for the gauge potential one representative on each gauge orbit, that is, by considering the quotient of the space of gauge connections with respect to the gauge group. Gribov ambiguity amounts to the fact that there are in general different field configurations which obey the same gauge-fixing condition, but which are related by a gauge transformation, i.e., they are on the same gauge orbit. As first shown by Singer \cite{Singer1978} and independently by Narasimhan and Ramadas \cite{Narasimhan1979}, this occurrence can be given a precise mathematical characterisation in terms of topological non-triviality of $\mathcal{G}$, the pertinent gauge group involved. On this basis,  QED is singled out, with the relevant homotopy group $\Pi_1(\widehat{U}(1))$ being trivial. For Quantum Chromodynamics (QCD), however, Gribov copies pose a real threat to the accurate description of the low-energy regime of the theory,  the gauge group of SU(N) Yang-Mills theories being nontrivial for $N\ge 2$. Over the last few decades, several solutions have been proposed in order to restrict the functional integral to the region in the space of  connections where only one representative for each gauge orbit is present, the so-called  first Gribov region. The most notable efforts are the Gribov-Zwanziger approach \cite{Zwanziger:1989mf} and its refined version \cite{Dudal:2008sp}. 

In \cite{Canfora2016, Kurkov2018} the problem has been   addressed  in the context of gauge theories on noncommutative space-time. It has been shown that, differently from the commutative case, noncommutative QED (NCQED) exhibits Gribov ambiguity and  gauge configurations have been found, with an infinite number of copies. The goal of the present work is to investigate the characterisation of the  first Gribov region.  Following the same approach as in non-Abelian gauge theories \cite{Dudal:2008sp}, the ensuing constraint on the physical space of connections is implemented  by requiring that  the ghost propagator does not develop poles. This implies a modification of the photon propagator. 

The paper is organised as follows. In section \ref{gaugeth},  we shortly review the problem in the framework of standard gauge theory. We then recall in section \ref{NCQED} the formulation of QED in the noncommutative setting with Moyal type noncommutativity and review the derivation of the equation for Gribov copies introduced in \cite{Canfora2016, Kurkov2018}. In section \ref{Gribovregion}, we apply the techniques developed in \cite{Dudal:2008sp} to characterise the first Gribov region for NCQED. In section \ref{five}, we evaluate the no-pole condition in the Landau gauge and find the Gribov region function $V(\Omega)$. Finally, we calculate the photon propagator in noncommutative momentum space. We conclude with some remarks and future perspectives.

\section{Gribov ambiguity in Yang-Mills}\label{gaugeth} 

Let $M=\R^4$ be the space-time  manifold and let us define the group of gauge transformations,   $\mathcal{G}=\widehat{U}(N)$, as the smooth maps $ {g:M\rightarrow U(N)}$,  with boundary condition $  g(x) \rightarrow 1$ as $ |{ x}| \rightarrow \infty$.

 A pure gauge  theory of fundamental interactions, without matter fields, is a theory where the dynamical fields are the gauge connections,  locally represented by Lie algebra valued one-forms $A$, with   $F= dA+ A\wedge A$ the local curvature two-form. When $M$ is the Euclidean space-time, the classical action describing the dynamics is 
\be
S=   \frac{1}{4}\int F^a_{\mu\nu} {F^{\mu\nu}}^a \dd^4 x \label{act}
\ee
where  $F= F^a_{\mu\nu} \tau_a dx^\mu\wedge dx^\nu$  and $\tau_a$ are the generators of the Lie algebra. The quadratic, free, sector of the theory, upon integrating by parts with suitable boundary conditions, can be written as 
\be
S= \frac{1}{2}\int \dd^n x\, {\int \dd^n y\,} A^a_\mu(x) \Delta_{ab}^{\mu\nu} (x,y) A^b_\nu(y), \label{action}
\ee
with
\be 
\Delta_{ab}^{\mu\nu} (x,y)= (-\partial^2\delta^{\mu\nu} +\del^\mu\del^\nu) \delta^{(4)} (x-y)\delta_{ab} \label{Mop}
\ee
and   
\be
Z[J]= \int_{\mathcal{A} }  [d \mu(\mathcal{A})] \exp\left(-\frac{1}{2}(S[A]+S_I[A,J])\right), \label{zeta}
 \ee
 the generating functional and  $\mathcal{A}$ the space of gauge connections.  If  the operator $\Delta_{ab}^{\mu\nu}$ were invertible, as it is the case for scalar theories,  the Gaussian integral in \eqn{zeta} could be {formally} performed to give:
\be 
Z[J] = \left(\det \Delta\right)^{-\frac{1}{2}} \exp\left( \frac{1}{2} \int J \Delta^{-1} J\right) 
\ee
with $\Delta^{-1}$ the Euclidean propagator, but this is not the case for gauge theories unless we manage to integrate over equivalence classes of gauge connections. Indeed, because of gauge invariance of the free action under 
$
A \rightarrow A^g= g A g^{-1} + dg g^{-1},  
$
field configurations of the form  $dg g^{-1}$ (so-called pure gauge terms) are in the kernel of $\Delta_{ab}^{\mu\nu}$, such that
\be
\Delta^{\mu\nu} \del_\nu g g^{-1}= 0, \label{zeromodes}
\ee
showing that the operator \eqn{Mop} has eigenvectors with zero eigenvalues (so-called zero modes).  To restrict the functional integration to gauge inequivalent potentials, one has to limit the functional integral \eqn{zeta} to the quotient space $\mathcal{B}= \mathcal{A}/\mathcal{G}.$
  Mathematically, this amounts to choosing a surface $\Sigma_{f} \subset \mathcal{A}$  which possibly intersects the gauge orbits only once.
The choice of $\Sigma_f$ is  physically rephrased as a {gauge fixing},  $f(A)= h$, for some chosen  functions $f, h$.  Within the standard quantization procedure of gauge theories, this is  achieved through the  introduction of the Faddeev-Popov determinant, which is only unambiguous when no topological obstructions arise.  But precisely the latter is at the origin of Gribov ambiguities. Therefore, we shortly summarise the standard procedure and the related topological issues.

The kinematical configuration space $\mathcal{A}$  is usually assumed to be  globally equivalent to the  product  $ \mathcal{B} \times \mathcal{G}$. In such a case, one has for the integration measure
\be \label{measure}
 [d \mu(\mathcal{A})]= [d\mu(\mathcal{B})]  ~ [d\mu (\mathcal{G}) ]= [d\mu(\mathcal{B})]  ~[d\alpha ],
 \ee
for gauge transformations close to the identity, $g(x) \simeq {\mathbf 1}+ \alpha^a (x) \tau_a$.  By    performing  a change of variables  $ [d\alpha ]\rightarrow  [d f(A) ]$, with the insertion of   the Jacobian   
\be
 \rm{Det} \frac{\delta f^{a}(x)}{\delta \alpha^{b}(y)}  \equiv \rm{Det}  \Delta (x,y) \label{FPdet},
\ee
one arrives at 
\be
[d \mu(\mathcal{A})] \rm{Det} \Delta  =  [{ [d\mu(\mathcal{B})] }  [d\alpha ]  \rm{Det} \Delta   =  { [d\mu(\mathcal{B})] }\; [d f ].
\ee
And, finally, by integrating over $[df]$ with the insertion of a delta function $\delta(f(A) -h(x))$ which implements the gauge choice, one obtains the measure on the quotient space:  
\be
  [d \mu(\mathcal{A})] \; \rm{Det} \Delta\;\; \delta(f(A) -h(x))={ [d\mu(\mathcal{B})] }. \label{measuregauged}
  \ee
  The Jacobian in \eqn{FPdet} is the known Faddeev-Popov determinant. 

However, the gauge fixing described above is not enough to remove unphysical degrees of freedom if the theory is non-Abelian. 
Indeed, let us consider the gauge orbit 
\be
A^g =g  A g^{-1} + d g g^{-1}  \simeq  A+ D\alpha,
\ee
with here  $ - (\del_\mu D^\mu) \delta^{(4)}(x-y) \delta^{ab} $ is the FP operator for this choice of gauge fixing and $D\alpha=  d\alpha+ \alpha^a\wedge A^b [\tau_a, \tau_b]$. 
 The gauge fixing condition  $\del^\mu A^g _\mu= 0$  yields the so-called  {\it equation of copies} \cite{Gribov1978}
 \be
 \del_\mu D^\mu \alpha= 0 ,\label{copies}
 \ee
 which may have nontrivial solutions, whenever the gauge group is non-Abelian, yielding to Gribov ambiguities.\footnote{\label{footnote3} In the Abelian case we only have trivial solutions, if we further assume that $\lim_{x\rightarrow\infty} \alpha(x)=0$.} 
The  problem can be traced back to the topological non-equivalence of the kinematical space of connections ${\mathcal A}$ and the physical space ${\mathcal B}$, \cite{Singer1978}, \cite{Narasimhan1979}. Indeed \eqn{measuregauged} is only valid if $\mathcal{A}= \mathcal{B} \times \mathcal{G}$ globally, which, for $\mathcal{G}=\widehat{U}(N)$ is only true for $N=1$  namely for QED (see for example \cite{Nair2005} for a clear pedagogical discussion). 

\section{The Gribov copies equation for  NC QED }\label{NCQED}
For each two functions $f, g$ defined on Euclidean space-time $\R^4$,  the
noncommutative Moyal star product $f\star g$ \cite{Moyal:1949sk} can be given the following asymptotic expansion 
\begin{equation}
(f\star g)(x)=f(x)\exp \left\{ \frac{i}{2}\,\Theta ^{\rho \sigma }\overset{%
\leftarrow }{\partial _{\rho }}\overset{\rightarrow }{\partial _{\sigma }}%
\right\} g(x),  \label{star}
\end{equation}%
with $\rho ,\sigma =1,..,4$.
 The antisymmetric matrix $\Theta $, has the following nonzero components 
\begin{equation}
\Theta _{1,2}=-\Theta _{2,1}=\Theta_1 ,\,\,\Theta _{3,4}=-\Theta _{4,3}=\Theta_2
 ,  \label{nctheta}
\end{equation}%
where $\Theta_i $ are real deformation parameters, in principle all different from each other, 
characterizing noncommutativity. For simplicity, we perform a rescaling in order to make all parameters equal. When $\Theta_i \rightarrow 0$, the star
product tends to the standard commutative point-wise product of $f$
and $g$. 
\subsection{Gauge transformations}
 Gauge theories  with gauge group ${\mathcal{G}}= {\rm Maps}(\R^4, U(N))$ 
are modified in the noncommutative setting by replacing the point-wise product with the non-local product~\eqn{star}. 
The elements of the gauge group,  $U_\star(x)$, are defined as star exponentials
\be
U_\star(x)=\exp_\star\left(i \alpha (x)^i T_i\right)= 1+ i \alpha^i
(x) T_i - \frac{1}{2}(\alpha^i\star \alpha^j)(x) T_i T_j + \ldots
\label{starexp}
\ee
with $T_i$ the Lie algebra generators of the structure group, $T_i \in \mathfrak{u}(N)$. 
The associated  matter fields transform under  deformed gauge transformations according to 
\be
\phi(x) \longrightarrow U_\star(x)\triangleright_\star \phi(x)=\exp_\star\left(i
\alpha^i (x) T_i\right) \triangleright_\star \phi (x). \label{gaugetransf}
\ee
 with $\triangleright_\star$ indicating simultaneously  the appropriate representation of the Lie algebra generators  {\it and} the $\star$ multiplication in space-time.  
 At the infinitesimal level, we have then
\be
\phi(x) \longrightarrow  \phi(x) +i (\alpha\triangleright_\star \phi)(x),
\label{infgaugetransf}
\ee
with
\be
(\alpha\triangleright_\star\phi)(x)=i \left(\alpha^j(x)\star (T_j\triangleright \phi)
\right)(x).
\ee
The gauge potential  transforms  as
\be
A_\mu \rightarrow A^{\prime}_{\mu} = U_\star \star A_{\mu} \star U_\star^{\dagger}
+i\, U_\star \star \partial_{\mu} U_\star^{\dagger}.  \label{gtransgen}
\end{equation}
Specialising to NCQED, the infinitesimal transformation  reads therefore
\begin{equation}
A_\mu \rightarrow A^{\prime}_{\mu}[\alpha] = A_{\mu} + D_{\mu}\alpha + \mathcal{O%
}(\alpha),  \label{gtransinfin}
\end{equation}
where the covariant derivative $D_{\mu}$, now  only due to space-time {noncommutativity}, is given by 

\begin{equation}
D_{\mu}\alpha = \partial_{\mu}\alpha + i \left(\alpha \star A_{\mu} - A_{\mu}\star \alpha\right)
\label{cd}
\end{equation}
and reduces to the standard Abelian form in the  commutative limit  $%
\Theta\rightarrow 0$.
 The field strength $F$  is given by
 \be
 F_{\mu\nu}= D_{\mu} A_\nu -D_{\nu} A_\mu
 \ee
 and transforms covariantly under star-gauge transformations.
 
It is precisely the occurrence of the covariant derivative which makes NCQED similar to non-Abelian gauge theories in commutative space-time, and, independently from the specific form of the gauge action $S[A]$ one deals with,\footnote{Besides the  natural candidate, which is obtained by replacing the point-wise product with the noncommutative one, i.e.
\be
S[A]= \frac{1}{4}\int F_{\mu\nu}\star F^{\mu\nu}
\ee
other proposals have been considered in the literature. See for example \cite{MVW14} and refs. therein.}  it is a meaningful question and a relevant issue for the quantization of the theory to investigate the existence of nontrivial solutions for the equation of Gribov copies. The latter is readily obtained by considering  $A_{\m}$ and $A'_{\m}$ satisfying  the same gauge condition, e.g. the Landau gauge,
\beq
\pa^{\m}A_{\m}=\pa^{\m}A'_{\m}=0.
\eeq
It follows that $\a$ yields a Gribov copy if  
\beq
-\pa^{\m}D_{\m}\a(x) &=& 0
\eeq
The problem has been addressed in \cite{Canfora2016, Kurkov2018} where the equation has been solved for a class of particularly simple potentials, and an  infinite number of copies, which cannot be discarded by boundary conditions, i.e. they are non-trivial,  has been found.  

 It is therefore clear that the standard  quantization procedure applied to NCQED does not provide a well-defined measure of integration over  gauge fields. In order to overcome the problem, we propose to apply the same techniques adopted for non-Abelian gauge theories, which have led to restricting the functional  integral  to an integration region free of copies: the Gribov region.

\section{Establishing the Gribov Region}\label{Gribovregion}
Once  the existence of copies has been established in terms of zero modes of the Faddeev-Popov operator, we need to characterise  the  region $\Omega$ in the space of connections where there are no ambiguities. The definition  which we adopt here for NCQED  was proposed by Gribov in 1978 for non-Abelian theories \cite{Gribov1978}. For the present case it reads:
\beq
\O \equiv \{A_{\m},\,\,\pa^{\m} A_{\m}=0,\,\, \Delta(x,y)>0\},
\eeq
where the Faddeev-Popov operator is explicitly  given by 
\beq\label{FPstar}
\Delta(x,y)=-\pa_{\m}D^{\m}\delta(x-y)=-\partial^2\delta(x-y)+i\lambda[A_{\m},\pa_{\m}\delta(x-y)]_{\star}
\eeq
and $ [f, g]_\star= f\star g -g\star f$. 

The boundary $\del \Omega$ represents  the first  Gribov horizon. For non-Abelian gauge theories it was shown \cite{DellAntonio:1989wae,DellAntonio:1991mms} that $\Omega$ is convex, bounded in all directions in the space of gauge connections  and that 
each gauge orbit passes at least once in  $\Omega$. The inverse of the FP 
operator, or equivalently the ghost propagator with external gauge field, 
$G(k, A)= \Delta^{-1}(k, A)$,  with $k$ the momentum variable, can 
be used to implement the restriction to $\Omega$ \cite{Gribov1978}. In the following we make the non-trivial assumption  that the results established 
in \cite{DellAntonio:1989wae,DellAntonio:1991mms} may be extended to NCQED, namely that
\begin{itemize}
\item
The Gribov region is bounded in every direction  (in the functional space 
of transverse gauge potentials).
\item  The Faddeev-Popov determinant changes sign at the Gribov horizon.
\item  Every gauge orbit passes inside the Gribov horizon.
\end{itemize}
Under these assumptions, we will adapt to NCQED   the procedure introduced in \cite{Capri:2012wx} to restrict  the
functional integration to the Gribov region. In the next subsection we provide some arguments to support these assumptions.


\subsection{ Properties of the NC Faddeev-Popov operator}

The FP operator is given by
\begin{align}
	\Delta(x,y, A) =-\partial_{\mu}D^{\mu}\delta(x-y) =-\partial^2\delta(x-y)+i\lambda[A_{\mu},\partial_{\mu}\delta(x-y)]_{\star} 
\end{align}	

\begin{itemize}
	\item \emph{\textbf{The Gribov region is limited in every direction}}
	
	First of all, note that  $\Delta(x,y, A)$ is hermitian in Euclidean space so that for real functions $f(x)$ and $g(x)$ one has 
	\begin{align}
	\int d^Dx \int d^D y\; f(x)\star \Delta(x,y, A) \star g(y) \in \mathds{R}
	\end{align}
	Now, suppose that $\Delta(x,y, A)$ develops a zero mode for some value of $A$. This must come from some negtive contribution from the term ${\cal M}(x,y, A) = i\lambda[A_{\mu},\partial_{\mu}\delta(x-y)]_{\star}$ because the term $\partial^2\delta(x-y)$ is positive. Therefore one must have for some value of $A$ and some function $\psi(x)$ that
	\begin{align}
		\int d^Dx \int d^D y\; \psi(x)\star {\cal M}(x,y, A) \star \psi(y) = a < 0
	\end{align}	
	Now introduce a real parameter $t$ multiplying $A$ so as to parameterize its amplitude isotropicaly and consider the expression 
		\begin{align}
			\int d^Dx \int d^D y\; \psi(x)\star \Delta(x,y, t A) \star \psi(y)  = \int d^Dx \int d^D y\; \psi(x)(-\partial^2) \psi(x)  + t a
		\end{align}	
	where we used that ${\cal M}(x,y, tA) = t {\cal M}(x,y, A)$. The first term is positive and since $a$ is negative, it is clear that this expression will become negative for some sufficiently high value of $t$. This means that the Gribov region, defined as the region in $A$ space such that $\Delta(x,y, A)$ is positive, will be eventually crossed for some finite value of $t$ and thus it must be limited in every direction.  
	
	\item \emph{\textbf{The Faddeev-Popov determinant changes sign at the Gribov horizon.}} This follows from the definition of the Gribov horizon and can be seen from the above remarks. As the amplitude of $A$ becomes large the $FP$ reaches a zero mode (the Gribov horizon) and afterward it becomes negative as we saw by raising the parameter $t$ in the proof above.
	
	\item \emph{\textbf{Every gauge orbit passes inside the Gribov horizon.}}
	In order to prove this, one would need a functional $f(A)$ such that under an infinitesimal  gauge transformation $\delta A_{\mu} = D_{\mu} \alpha$ its extrema are determined by the gauge condition
	\begin{align}
	\delta f(A) = 0 \Rightarrow \partial_{\mu} A_{\mu} = 0
	\end{align}	
	and its second variation is positive, thus making these extrema in fact the minima
	\begin{align}
		\delta^2 f(A) > 0 \Rightarrow \partial_{\mu}D_{\mu}  > 0
	\end{align}	
	 The argument then go on to show that for every gauge orbit, as one moves along the orbit, the functional $f(A)$ will eventually attain its absolute minimum. But since the condition for the minimum is exactly the condition that defines the Gribov region, $\partial_{\mu}D_{\mu}  > 0$, it would follow that every gauge orbit will pass through the Gribov region. For the usual non-abelian case, the functional is $f(A) = \frac 12 \int d^D x A^a_{\mu}A^a_{\mu}$.  
	
 We have verified that the same functional works for the NC case, by using the cyclicity of the product. Given 
 \begin{eqnarray}
 	f(A)= \int d^Dx\, A_\mu \star A_\mu = \int d^Dx\, A_\mu  A_\mu .
 \end{eqnarray}
The first variation is
\begin{eqnarray}
	0= \delta f(A) &=&  \int d^Dx\, A_\mu \star \delta A_\mu =  \int d^Dx\, A_\mu \star (\partial_\mu \alpha + i [ \alpha, A_\mu]_\star)= \int d^Dx\, A_\mu \star \partial_\mu \alpha \nonumber\\
	&=& \int d^Dx\, A_\mu\cdot  \partial_\mu \alpha    =\int d^Dx\, \partial _\mu A_\mu \cdot \alpha=\int d^Dx\, \partial _\mu A_\mu \star \alpha, 
\end{eqnarray}
which implies $\partial _\mu A_\mu= 0$.  Here  we have used the cyclicity of Moyal  product under the integral  ($\int a\star b\star c$ is cyclic) and $\int a\star b= \int b\star a$. Notice that the last property is in general not true for other star products. Then the second variation gives
\begin{eqnarray}
	0<\delta^2 f(A)= \int dx \delta \partial_\mu  A_\mu \star \alpha= \int \partial_\mu D_\mu \alpha \star \alpha = \int \partial_\mu D_\mu \alpha \cdot \alpha
\end{eqnarray}
which, for arbitrary $\alpha$ implies the positivity of  $\partial_\mu D_\mu$, namely the condition defining the Gribov region.  As mentioned above, to provide a complete proof one would need the show that the functional $f(A)$ attains its absolute minimum along every orbit in the NC case as well. This is a reasonable expectation but to pursue such a formal proof is beyond the scope of this work. 
\end{itemize}

\section{NCQED no-pole condition}\label{five}

 The requirement that the FP operator has no zero eigenvalues is implemented in the standard approach \cite{Gribov1978}  in terms of the request that the 
inverse of the operator doesn't have poles: this is the  so-called {\it no pole condition}. The latter is seen to restrict  
the region of integration to  the Gribov region  by introducing   a factor $V(\O)$ in the generating functional,  
\be\label{fungen}
Z(A,c,\bar c)=\int_{\O}\,\, [dA][d\bc][dc]\,V(\O)\, \d(\pa\cdot A) 
\exp\left[-S[A]-\int\,dx\,dy \,\bc(x)\star \Delta(x,y) \star c(y) \right]
\ee
with $\Delta(x,y)$ given in \eqref{FPstar}. We recall that we are in the Euclidean setting.
We work in the Landau gauge, $\d(\pa\cdot A)$. To determine $V(\O)$ we use the relationship between the ghost sector of the theory and the FP operator, which emerges from calculating the exact ghost propagator
\be
\langle\bar c(p) \,c (-p)\rangle= \frac{\delta}{\delta J_c(y)}\frac{\delta}{\delta J_{\bar c}(x)} Z[J]
\ee
with 
\beq
Z[J]&=&\int_{\O}\,\, [dA][d\bc][dc]\,V(\O)\, \d(\pa_{\m}A_{\m}) \\
&\times & \exp\left[-S[A]-\int\,dx\,dy \, \bc(x)\star \Delta(x,y) \star c(y) +\int \, dx\, \left(J_c(x) \star c(x)+ \bar c(x) \star J(x)\right) \nn\right].
\eeq
Thanks to the  cyclicity and closure properties of the Moyal product\footnote{They amount respectively  to the fact that $\int f_1\star ...\star f_n= \int f_n\star f_1\star...$ and $\int f\star g= f \cdot g$. }\label{cycl} we may perform the Gaussian integral over ghost fields in standard fashion, obtaining
\beq
Z(J)=C \det (\Delta) \exp\left[-\int d^d x d^d y J_c(x)\star \Delta^{-1}(x,y)\star J_{\bc}(y)\right]
\eeq
so that 
\be
\langle\bar c(p) \,c (-p)\rangle= \int [d A] V(\Omega) \delta(\del^\mu A_\mu) \det(-\del^\mu D_\mu)  \Delta^{-1}(x,y) \exp(-S[A])
\ee
with  $\Delta^{-1}(x,y)$ the ghost propagator at first order in $\hbar$.  
According to \cite{Capri:2012wx} it is convenient to interpret $\Delta(x,y)$ as the space representation of an abstract operator $\Delta$ 
\be
\Delta(x,y)= \langle x|\Delta|y \rangle.
\ee
Inserting completeness relations  $\int d^dx\, |x\rangle\langle x|=1$, we may obtain   $\Delta(x,y)$  in momentum space
\be
\Delta(p,q)= \int d^dx\int d^dy \langle p|x\rangle\star\langle x|\Delta |y\rangle\star \langle y|q\rangle=\int d^d x \int d^d y \,e^{-ipx}\star\Delta(x,y)\star e^{iqy}.
\ee
The explicit calculation of star products, performed in  App. \ref{appA} yields 
\be\label{mpq}
\Delta(p,q)=
q^2\delta(p-q)+2i\lambda \sin\left(\frac{1}{2}\Th_{\r\s}p_{\r}q_{\s}\right)\tilde A_{\m}(p-q) q_{\m}.
\ee  
where
\beqa
\langle x|p\rangle&=& \frac{e^{ipx}}{(2\pi)^{d/2}}\nn\\
\tilde A_{\m}(p-q)&=& \int \frac{d^d x }{(2\pi)^d} e^{-i(p-q)x}\star A_\mu(x)= \int \frac{d^d x }{(2\pi)^d} \,e^{-i(p-q)x} A_\mu(x)
\eeqa

Upon introducing the matrix notation 
\beq
\mathbbm{1}_{pq}&=&\braket{p\vert \mathbbm{1}\vert q}=\delta(p-q),\\
\mathbbm{A}_{pq}&=&\braket{p\vert \mathbbm{A}\vert q}=-2i\sin\left(\frac{1}{2}\Th_{\r\s}p_{\r}q_{\s}\right)\tilde A_{\m}(p-q)\frac{q_{\m}}{q^2},
\eeq
we may rewrite Eq. \eqn{mpq} as
\beq
\Delta_{pq}=q^2(\mathbbm{1}_{pq}-\l\mathbbm{A}_{pq}).
\eeq

The no-pole condition works as a bound on the allowed amplitudes of the gauge field configurations and  implies that inverse of this expression makes sense. Then, the inverse of the operator $\Delta$ can be written as
\beq
\Delta_{pq}^{-1}=\frac{1}{q^2(\mathbbm{1}_{pq}-\l\mathbbm{A}_{pq})}=\frac{1}{q^2}\sum_{n=0}^{\infty}[(\l\mathbbm{A})^n]_{pq}.\label{eq8}
\eeq
We are interested in the poles of this expression, it is then convenient to study the corresponding normalized trace 
\beq
\frac{1}{V}\left.\Delta_{pq}^{-1}\right\vert_{p=q=k}\equiv \frac{1}{k^2}\left(1 +  \s(k,A)\right).\label{formfactor}
\eeq
with $V=\int \frac{d^dx}{(2\pi)^d}$ the infinite volume factor. This equation defines the so-called form factor $\s(k,A)$, which encapsulates all non-trivial information about the pole structure of the full ghost propagator, in fact, if it vanishes we recover the free propagator expression. In order to implement the no-pole condition one has to integrate over all gauge field configurations leading to 
\beq\label{1PI}
\Delta^{-1}_{kk}(k) =\langle \Delta^{-1}_{kk}(k, A) \rangle_{conn} = \frac{V}{k^2} \left(1 + \langle \sigma(k,A) \rangle_{conn} \right) = \frac{V}{k^2} \frac{1}{1- \langle \sigma(k, A)\rangle_{1PI} }
\eeq
where $\langle \cdots \rangle_{conn}$ is the sum of  all connected diagrams and $\langle \cdots \rangle_{1PI}$ is the sum of all one-particle-irreducible diagrams (see \cite{Capri:2012wx} for details). The no-pole condition is the requirement that the ghost propagator does not develop a pole for any value of $k\neq 0$. The evaluation of the exact no-pole condition thus amounts to the computation of the exact value $\langle \sigma(k, A)\rangle_{1PI}$. This is of course a very difficult problem ( see the discussion in Appendix (\ref{appALLORDERS})) and we will study here an approximate solution by considering only the first non-trivial contribution. 
From expression \eqref{eq8}, we get
\beq\label{M-1}
k^2\Delta^{-1}_{kk}(k, A)=\mathbbm{1}_{kk}+g\mathbbm{A}_{kk}+\l^2\mathbbm{A}^2_{kk}+\mc{O}(\l^3).
\eeq
Then, we can write $\sigma(k,A)$ as
\beq
\s(k,A)&=&\frac{1}{V }(\mathbbm{1}_{kk}+\lambda \mathbbm{A}_{kk}+\l^2\mathbbm{A}^2_{kk}+ \mc{O}(\l^3))-1.\\
&=& \frac{\lambda^2}{V} \int \frac{d^d p}{(2\pi)^{d}} \int \frac{d^d q}{(2\pi)^{d}}\mathbbm{A}_{kp}\left(  q^2 \Delta_{pq}^{-1}  \right)\mathbbm{A}_{qk}
\label{eq540}
\eeq
where we used \eqref{eq8} and the relations
\beq
\mathbbm{1}_{kk}&=&V,\\
\mathbbm{A}_{kk}&=&-2\sin(0)A_{\m}(0)\frac{ik_{\m}}{k^2}=0,
\eeq
The first nontrivial contribution to $\s(k,A)$ is thus obtained approximating the  ghost propagator inside the integral \eqref{eq8} by its tree-level expression, that is  $q^2 \Delta_{pq}^{-1} \approx \mathbbm{1}_{pq}$. Therefore, noting that
\beq
\mathbbm{A}^{2}_{kk}&=&\int \frac{d^d q}{(2\pi)^{d}}\, \braket{k\vert \mathbbm{A}\vert q}\braket{q\vert \mathbbm{A}\vert k}\nn\\
&=&-4\int \frac{d^d q}{(2\pi)^{d}}\sin\left(\frac{1}{2}\Th_{\r\s}k_{\r}q_{\s}\right)\tilde A_{\m}(k-q)\frac{q_{\m}}{q^2}\sin\left(\frac{1}{2}\Th_{\r\s}q_{\r}k_{\s}\right)\tilde A_{\n}(q-k)\frac{k_{\n}}{k^2}\label{eq654}
\eeq
equation \eqref{eq540} becomes
\beq
\s(k,A,\Th)=-\frac{4\l^2}{V}\int \frac{d^d q}{(2\pi)^{d}}\sin\left(\frac{1}{2}\Th_{\r\s}k_{\r}q_{\s}\right)\tilde A_{\m}(k-q)\frac{q_{\m}}{q^2}\sin\left(\frac{1}{2}\Th_{\r\s}q_{\r}k_{\s}\right)\tilde A_{\n}(q-k)\frac{k_{\n}}{k^2}
\eeq
where the $\Theta$ dependence was highlighted in the argument of $\s$. Considering the Landau gauge $q_{\m}A_{\m}(k-q)=k_{\m}A_{\m}(k-q)$ and changing $q\rightarrow q+k$, we obtain, 
\beq
\s(k,A,\Th)&=&\frac{4\l^2}{V}\frac{k_{\m}k_{\n}}{k^2} \int \frac{d^d q}{(2\pi)^{d}} \sin^2\left(\frac{1}{2}\Th_{\r\s}q_{\r}k_{\s}\right)\tilde A_{\m}(-q)\frac{1}{(k+q)^2}\tilde A_{\n}(q).
\eeq
The latter may be further simplified by observing that, in the Landau gauge, $\tilde A_\mu(-q) \tilde A_\nu(q)$ is transversal, 
\be
\tilde A_\mu(-q) \tilde A_\nu(q)=\omega(A)\left(\delta_{\mu\nu}-\frac{q_\mu q_\nu}{q^2}\right).
\ee
Multiplying by $\delta_{\mu\nu}$ the factor $\omega(A)$ is found to be    $\omega(A)= \frac{1}{d-1}|\tilde A|^2$. 
Thus we obtain
\beq
\s(k, A,\Th)=\frac{4\l^2}{V}\frac{1}{d-1}\frac{k_{\mu}k_{\nu}}{k^2} \int \frac{d^d q}{(2\pi)^{d}} \sin^2\left(\frac{1}{2}\Th_{\r\s}q_{\r}k_{\s}\right)\frac{\vert A_{\l}(q)\vert^2}{(k+q)^2}\left(1-\frac{q_{\mu}q_{\nu}}{q^2}\right)\label{nopole}.
\eeq

Observing that $\dfrac{\sin^2(x)}{x^2}\leq 1$ we have $\s(k,\Th)\leq I(k,\Th)$. The function $I(k,\Th)$ is defined as 
\beq
I(k, A,\Th) &=&\frac{\l^2}{V}\frac{\Th_{\r\s}\Th_{\o\l}}{d-1}\frac{k_{\mu}k_{\nu}k_{\s}k_{\l}}{k^4} \int \frac{d^d q}{(2\pi)^{d}} \vert \tilde A_{\l}(q)\vert^2\frac{k^2 q^2}{(k+q)^2}\left(\delta_{\m\n}-\frac{q_{\mu}q_{\nu}}{q^2}\right)\frac{q_{\r}q_{\o}}{q^2}\\
&=&\frac{k_{\m}k_{\n}k_{\l}k_{\s}}{k^4} \mc{I}_{\m\n\l\s}(k,\Th),\label{nopolesmall}
\eeq
with
\beq
\mc{I}_{\m\n\l\s}(k,A,\Th )&=&\frac{\l^2}{V}\frac{\Th_{\r\s}\Th_{\o\l}}{d-1} \int \frac{d^d q}{(2\pi)^{d}} \vert \tilde A_{\l}(q)\vert^2\frac{k^2 q^2}{(k+q)^2}\left(\delta_{\m\n}-\frac{q_{\mu}q_{\nu}}{q^2}\right)\frac{q_{\r}q_{\o}}{q^2}\label{functionnopolesmall}.
\eeq
We are interested in finding an upper-bound for the latter, so to have a refined no-pole condition $I(k,\Th)<1$.
In Appendix \eqref{apa} and \eqref{apb}  we perform an explicit analysis in $d=2$ and $d=4$ and show that $I(k,A,\Theta)$ is an increasing function of $k$, with the maximum value  attained at $k\rightarrow \infty$. In what follows we shall therefore assume $d=2,4$, although  the analysis could  be probably extended to other dimensions. 
Because of that, the  no-pole condition is certainly satisfied if we impose
 \beq
 I_{max}(A,\Th )<1,
 \eeq
 with  
 \beq
 I_{max}(A,\Th )=\frac{k_{\m}k_{\n}k_{\l}k_{\s}}{k^4} \mc{I}_{\m\n\l\s}(\infty,A,\Th ).\label{lema}
 \eeq
Taking the limit $k^2\rightarrow \infty$ in \eqref{functionnopolesmall}, we obtain
\beq
\mc{I}_{\m\n\l\s}(\infty, A,\Th)&=&\frac{\l^2}{V}\frac{\Th_{\r\s}\Th_{\o\l}}{d-1} \int \frac{d^d q}{(2\pi)^{d}} \vert \tilde A_{\l}(q)\vert^2\left(\delta_{\m\n}q_{\r}q_{\o}-\frac{q_{\mu}q_{\nu}q_{\r}q_{\o}}{q^2}\right). \label{fff}
\eeq
Using the following properties of regularized momentum integrals \cite{Anacleto:2014aha}
\beq
\int \frac{d^d q}{(2\pi)^{d}} f(q^2) q_{\r}q_{\o}&=&\frac{\delta_{\r\o}}{d}\int \frac{d^d q}{(2\pi)^{d}} \, q^2 f(q^2),\label{id1}\\
\int \frac{d^d q}{(2\pi)^{d}} f(q^2) q_{\o}q_{\r}q_{\m}q_{\n}&=&\frac{\delta_{\m\n}\delta_{\o\r}+\delta_{\m\r}\delta_{\n\o}+\delta_{\m\o}\delta_{\r\n}}{d(d+2)}\int \frac{d^d q}{(2\pi)^{d}} \, q^4 f(q^2),
\eeq
we rewrite \eqref{fff} as
\beq
\mc{I}_{\m\n\l\s}(\infty, A,\Th )&=&\frac{\l^2}{V}\Th_{\r\s}\Th_{\o\l} \left(\frac{(d+1)\delta_{\m\n}\delta_{\r\o}-\delta_{\m\r}\delta_{\n\o}-\delta_{\m\o}\delta_{\r\n}}{d(d+2)(d-1)}\right)\int \frac{d^d q}{(2\pi)^{d}}\, q^2\vert \tilde A_{\l}(q)\vert^2\nonumber\\
&=&\frac{\l^2}{V}\left(\frac{(d+1)\delta_{\m\n}\Theta^2_{\s\l}-\Theta_{\m\s}\Theta_{\n\l}-\Theta_{\m\l}\Theta_{\n\s}}{d(d+2)(d-1)}\right) \int \frac{d^d q}{(2\pi)^{d}}\, q^2\vert \tilde A_{\l}(q)\vert^2. \label{eq764}
\eeq
 Substituing \eqref{eq764} in \eqref{lema}, we conclude that
\beq\label{Imax} 
I_{max}( A,\Th )&=&\frac{\l^2}{V}\frac{k_{\m}k_{\n}k_{\l}k_{\s}}{k^4} \left(\frac{(d+1)\delta_{\m\n}\Theta^2_{\s\l}-\Theta_{\m\s}\Theta_{\n\l}-\Theta_{\m\l}\Theta_{\n\s}}{d(d+2)(d-1)}\right) \int \frac{d^d q}{(2\pi)^{d}} \, q^2 \vert \tilde A_{\l}(q)\vert^2\nonumber\\
&=&\frac{\l^2}{V} \left(\frac{(d+1)\Theta^2}{d(d+2)(d-1)}\right) \int \frac{d^d q}{(2\pi)^{d}} \, q^2 \vert \tilde A_{\l}(q)\vert^2 ,
\eeq
where we have used  $k_{\s}k_{\l}\Theta^2_{\s\l}=k^2\Th^2$ and $k_{\m}k_{\s}\Th_{\m\s}=0$. 

Thanks to the latter, we may  estimate  the first Gribov region and exhibit its dependence on $\Theta$. Indeed, since $\sigma(k,A,\Theta)$ is always smaller or equal than $I(k,A,\Theta)$, which we have shown to be an increasing function of $k$, we may choose in Eq. \eqn{fungen}
\begin{equation}
V(\Omega)=\vartheta(1-I_{max}(A, \Theta)).
\end{equation} 
or using the Heaviside function integral form,
\beq
V(\Omega)= \int_{-\infty +\epsilon}^{+\infty +\epsilon}\frac{d\tau}{2\pi i \tau}e^{\tau(1-I_{max}(\Th))}.
\eeq
We can insert this into the path integral \eqref{action} which takes the form
\beq\label{Zfunctional}
Z(J)=C\int \frac{d\t}{2\pi i \t}e^{\t}\int[dA]e^{-\t I_{max}(A,\Theta)-\int d^dx\,\,\frac{1}{4}F_{\m\n}\star F_{\m\n} + \frac{1}{2\a}b \star\pa_{\m}A_{\m} +A_{\m}\star J_{\m}}. 
\eeq
The tree-level photon propagator can be read from the quadractic part and is given by 
\beq
\Delta_{\m\n}(k^2)&=&\left((1+\gamma)k^2 \d^{\m\n}+\left(\frac{1}{\a}-1\right)k_{\m}k_{\n}\right),\,\,\gamma \equiv\frac{2(d+1)}{d(d+2)(d-1)}\frac{\tau\,\Th^2 \l^2}{V}.
\eeq
Therefore, the transverse photon propagator is given by
\beq
\langle \tilde A_{\m}(-k)\tilde A_{\n}(k)\rangle &=&\frac{1}{(1+\gamma)k^2}\left(\delta_{\m\n}-{k_{\m}k_{\n}\over k^2}\right). 
\eeq
We see that, {\it at least within Moyal type non-commutativity}\footnote{Indeed, if the star product is not closed, already the quadratic part of the action, consisting of $\int (\del_\mu A_\nu-\del_\nu A_\mu) \star  \del_\mu A_\nu-\del_\nu A_\mu)$ is deformed. Namely, the $\star$ product cannot be removed, resulting in a modification of $\Delta_{\mu\nu}(k^2)$.} there is no qualitative modification of the photon propagator, which is changed only by a scale factor that can be absorbed by a $\Theta$ dependent renormalization of the gauge field. So the effect of Gribov copies is not felt and, in particular, there is no confining phase due to Gribov effects. It is a situation very similar to the case of the ${\cal N} = 4$ Super Yang-Mills, where Gribov effects are suppressed and no scale is generated \cite{Capri:2014tta}, thus maintaining the conformal invariance of the theory even when Gribov copies are taken into account. In the present case, the scale associated with the Gribov region is the noncommutative parameter $\Theta$, which is not dynamically generated but is already present as a fundamental scale of the theory from the start.  

 This is to be contrasted with the original (commutative space) non-abelian Gribov analysis \cite{Gribov1978}, where the modification of the quadratic part of the action coming from the form factor is $\sim \frac{\vert \tilde A_{\l}(q)\vert^2}{q^2}$ and it leads to an effective gluon propagator of the form  $\sim \frac{q^2}{q^4 + \kappa^4}$, where $\kappa$ is the Gribov parameter determining the scale associated with the Gribov horizon. This scale is dynamically generated and can be computed by a gap equation,
 \beq
 \int \frac{d^d q}{(2\pi)^{d}} \frac{1}{q^4 + \kappa^4} = 1
 \eeq
 whereas in the present case, the modification of the quadratic part induced by $I_{max}( A,\Th )$, \eqref{Imax},  is of the form $\sim \gamma q^2\vert \tilde A_{\l}(q)\vert^2$, with $\gamma$ proportional to $\Theta^2$.  This produces an ill-defined gap equation. In fact, the gap equation is obtained by extremising the vacuum energy functional, ${\cal E}(\tau) = - \ln Z$, with respect to $\tau$. So, after integrating over the gauge fields in \eqref{Zfunctional} considering only the quadratic part, we would need to find the extremum of
 \beq
 f(\tau) = -\ln \tau + \tau + \int \frac{d^d q}{(2\pi)^{d}} \ln \left((1+\gamma )q^2\right)
 \eeq
 The integral in this expression is ill-defined and in fact it is zero under dimensional regularization. Considering further that we eventually want to take the thermodynamic limit $V\rightarrow \infty$, keeping $\gamma$ finite (which means take $\tau \rightarrow \infty$), we see that this equation provides no consistent solution in this limit. Therefore, within the approximations considered, there is no dynamically generated scale  associated with the Gribov region in  NCQED. We conclude that the scale associated with the Gribov parameter in the non-abelian setting  is replaced by the nondynamical noncommutative scale here. We can observe that in the commutative limit,  when $\Theta \to 0$, all Gribov features are removed and the theory returns to  standard QED. 

As a final, important, remark, notice that our analysis is not exact. Besides the approximation of working only to leading nontrivial order in the form factor $\sigma(k, A, \Theta)$, we also consider a stronger no-pole condition than the needed one, since $I(k, A, \Theta)$ is an upper bound for $\sigma(k, A, \Theta)$. The result could be made more precise by maximising directly \eqref{nopole}. There are however technical problems which we haven't succeeded to solve for the moment.

\section{Conclusion}

We have analysed  the no-pole condition for NCQED with Moyal type non-commutativity.  By exploiting  the formal analogy of the latter with non-Abelian gauge theories, we have restricted  the effective action to the Gribov region by introducing a constraining factor, $V(\Omega)$, which is obtained through the  ghost propagator in a fashion similar to $SU(N)$ gauge theories. As already stressed, our result is to be intended as a first  estimate of the Gribov region, not only because we work to leading nontrivial order in the form factor $\sigma(k,A,\Theta)$, but also because the request that   $I(k,\Th)<1$ is a stronger no-pole condition than the needed one, $I(k,\Th)$  representing  an upper bound for $\sigma(k, A, \Th)$. This is due  to technical difficulties in maximising $\sigma(k,\Th)$ in a $k$-invariant way. Still, this upper bound  guarantees the absence of poles for the ghost propagation. Also, it permits the calculation of a non-commutative transversal photon propagator proportional to $\dfrac{1}{(1+\gamma)k^2}$, which results to be  dependent on $\Th^2$, the squared non-commutative parameter.  We therefore conclude that, in the approximation chosen,  there is no qualitative modification of the photon propagator, it being only  changed  by a scale factor that can be absorbed by a $\Theta$ dependent renormalisation of the gauge field. Let us stress, however, that this result has only been proven for the Moyal star product. Already mild modifications of the latter, such as the Wick-Voros product, would introduce a momentum dependent weight which could modify the behaviour of the propagator (see  for example \cite{Galluccio:2008wk} for an application to scalar field theory). It would be interesting to analyse this aspect more in detail.  
Some immediate research questions can be investigated following these results. First, naturally suggested by the present analysis we would like to find a more precise solution of $\sigma(k,\Th)<1$  dependent on the non-commutative parameter. Another possible path is to extend the analysis beyond the original Gribov semiclassical treatment and construct the analogous of the full Gribov-Zwanziger action \cite{zwanziger1989action, Zwanziger:1989mf, zwanziger1993renormalizability} for the NCQED. Finally, for the reasons outlined above, it would be interesting to repeat the analysis for a different kind of non-commutativity.

\section*{Acknowledgments}
O.H. acknowledges support  by the São Paulo Research Foundation (FAPESP), by the grant 2019/26291-8.  M. S. Guimaraes is a level 2 CNPq researcher under the contract 310049/2020-2.
L. Rosa and P. Vitale are partially supported by INFN.

\begin{appendices}
\section{Faddeev-Popov operator in   momentum space}\label{appA}
We start writing $\Delta (x,y)$ in the momentum space as 
\beq
\Delta(p,q)&=&\int d^d x \int d^d y \,e^{-ipx}\star\Delta(x,y)\star e^{iqy}.
\eeq  
We can write that as
\beq
\Delta(p,q)&=&\int d^d x \int d^d 
y \,e^{-ipx}\star(-\partial^2\delta(x-y)+i\l[A_{\m},\pa_{\m}\delta(x-y)]_{\star})\star e^{iqy},\nn\\
&=&\int d^d x \int d^d 
y \,e^{-ipx}\star(-\partial^2\delta(x-y)+i\l A_{\m}(x)\pa_{\m}\star\delta(x-y)\nonumber\\&-&i\l\pa_{\m}\star\delta(x-y)\star A_{\m}(x))\star e^{iqy}.\label{faddeev}
\eeq  
By substituting the delta function in momentum space 
\beq
\delta(x-y)=\int \frac{dk}{2\pi ^{d}} e^{i k (x-y)}\label{deltafunction}
\eeq
we arrive at
\beq
\Delta(p,q) &=&\int d^d x \int d^d 
y\int \frac{dk}{(2\p)^{d}} \,e^{-ipx}\star\Bigl(k^2e^{i k (x-y)}-\l A_{\m}(x)\star k_{\m}e^{i k (x-y)}\nonumber\\&+&\l k_{\m}e^{i k (x-y)}\star A_{\m}(x)\Bigr)\star e^{iqy}.
\eeq
On using  Eq. \eqn{star} to perform the  star   product in the $x$ and $y$ variables we get as an intermediate step
\beq
\Delta(p,q) =\int d^d x \int d^d 
y\int \frac{d^dk}{(2\p)^{d}} \,e^{-ipx}\star\Bigl(k^2e^{i k (x-y)+iqy}e^{\frac{i}{2}\Th_{\r\s}k_{\r}q_{\s}}\nonumber\\-\l A_{\m}(x)\star k_{\m}e^{i k (x-y)+iqy}e^{\frac{i}{2}\Th_{\r\s}k_{\r}q_{\s}}\Bigr)+\l k_{\m}e^{i k (x-y)-ipx}e^{\frac{i}{2}\Th_{\r\s}k_{\r}p_{\s}}\star A_{\m}(x)\star e^{iqy}\label{moyalfp}
\eeq
By computing the  first term in \eqref{moyalfp} and using the  closure of the star product (see footnote \ref{cycl}), we obtain
\beq
&&\int  d^d x \int d^d 
y\int \frac{d^dk}{(2\p)^{d}} \,e^{-ipx}\star k^2e^{i k (x-y)+iqy}e^{\frac{i}{2}\Th_{\r\s}k_{\r}q_{\s}}\nn\\
&=& \int  d^d x\int \frac{d^dk}{(2\p)^{d}} e^{-i p x} k^2 e^{ikx}\delta(q-k)e^{\frac{i}{2}\Th_{\r\s}k_{\r}q_{\s}}\nn\\
&=& \int  d^d x q^2  e^{-i p x} e^{iqx}=
q^2 \delta(q-p).\label{moyalfp1}
\eeq
From the second term in \eqref{moyalfp}, using the cyclicity and closure of the product
\beq
&-&\l\int d^d x \int d^d y\int \frac{d^dk}{(2\p)^{d}} \,k_{\m} e^{\frac{i}{2}\Th_{\r\s}k_{\r}q_{\s}}\, e^{-ipx}\star A_{\m}(x)\star e^{i k (x-y)+iqy}\nn\\
&=& -\l\int d^d x \int \frac{d^dk}{(2\p)^{d}} \,k_{\m}\delta(q-k) e^{\frac{i}{2}\Th_{\r\s}k_{\r}q_{\s}}\,e^{-ipx}\star A_{\m}(x)\star e^{ikx}\nn\\
&=& -\l\int d^d x \, q_{\m}\,e^{-ipx}\star A_{\m}(x)\star e^{iqx}\nn\\
&=& -\l\int d^d x \, q_{\m}\,e^{iqx}\star e^{-ipx}\star A_{\m}(x)\nn\\
&=& -\l\int d^d x \, q_{\m}\, e^{\frac{i}{2}\Th_{\r \s}q_\r p_\s} e^{i(q-p)x}\star A_{\m}(x)=-\l\int d^d x \, q_{\m}\, e^{\frac{i}{2}\Th_{\r \s}q_\r p_\s} e^{i(q-p)x} A_{\m}(x)\nn\\
%
&=& -\l q_{\m}A_{\m}(p-q) e^{\frac{i}{2}\Th_{\r \s}q_\r p_\s}.\label{moyalfp2}
\eeq
Following the same procedure in the third term in \eqref{moyalfp},
\beq
&&\l\int d^d x \int d^d y\int \frac{dk}{(2\p)^{d}} \, k_{\m}e^{\frac{i}{2}\Th_{\r\s}k_{\r}p_{\s}}\, e^{i k (x-y)-ipx}\star A_{\m}(x)\star e^{iqy}\nn\\
&=&\l q_{\m}A_{\m}(p-q) e^{\frac{i}{2}\Th_{\r \s}p_\r q_\s}\label{moyalfp3}
\eeq
Inserting \eqref{moyalfp1}, \eqref{moyalfp2} and \eqref{moyalfp3} in \eqref{moyalfp}, we prove that
\beq
\Delta(p,q) &=&q^2 \delta(p-q)+\l q_{\m}A_{\m}(p-q)\left(e^{\frac{i}{2}\Th_{\r \s}p_\r q_\s}-e^{-\frac{i}{2}\Th_{\r \s}p_\r q_\s}\right)\nn\\
&=&q^2 \delta(p-q)+2i\l q_{\m}A_{\m}(p-q)\sin\left(\frac{1}{2}\Th_{\r\s}p_{\r}q_{\s}\right).\label{deltapq}
\eeq
\section{Analysis of $\sigma(k, \Theta)$ in two dimensions}\label{apa}
In this Section we consider the function $\sigma(k, \Theta)$ in $d=2$ dimensions.  In polar coordinates of the plane, $(q, \alpha)$, $q^2= q^\mu q_\mu, \alpha\in [0, 2\pi]$,  Eq. \eqn{nopole} becomes 
\begin{equation}\label{sigma2}
\sigma(k, \Theta)=\frac{4 \lambda^2}{V}\int_0^\infty \frac{dq}{4\pi^2}\;q|A(q)|^2\int_0^{2\pi} d\alpha \frac{\sin^2{\left(\frac{1}{2}\Theta q k \sin{\alpha} 
\right)  } \sin^2{\alpha} }{(k-q)^2},
\end{equation}
where we have chosen $k_\mu = (0,k)$.
Observing  that $\frac{\sin^2x}{x^2}\leq1$, we have
\begin{eqnarray}
\sigma(k, \Theta) &\leq& I(k, \Theta)  =\frac{4 \lambda^2}{V}\int_0^\infty \frac{dq}{4\pi^2}\;q|A(q)|^2\int_0^{2\pi} d\alpha \frac{\left( {\frac{1}{2}\Theta q k \sin{\alpha}   } \right)^2 \sin^2{\alpha} }{(k-q)^2} \\
&& \nonumber\\
&=&\frac{\lambda^2}{4\pi V}\int_0^\infty dq\;q|A(q)|^2\frac{1}{4}\Theta^2 q^2 k^2 \left(
\frac{\left(3 k^2-q^2\right)}{4 k^4} \vartheta (k-q)-\frac{\left(k^2-3 q^2\right)}{4 q^4} \vartheta (q-k)
\right) \nonumber
\end{eqnarray}
where $\vartheta(x)$ is the Heaviside step function. We are interested in finding an upper-bound for the latter. Therefore we compute
\begin{eqnarray}
\frac{\partial I(k,\Th)}{\partial k}&=&\frac{ \lambda^2}{V}\Theta^2 \int_0^\infty dq\;q^3|A(q)|^2 \left(
\frac{k^2 \left(k^2-3 q^2\right) \delta (q-k)}{16 q^2}+\frac{q^2 \left(3 k^2-q^2\right)
   \delta (k-q)}{16 k^2} \right. \\
   &&\nonumber \\
   &&\left.+\frac{k^4 \left(3 q^2-2 k^2\right) \vartheta (q-k)+q^6 \theta
   (k-q)}{8 k^3 q^2}\right)  \nonumber \\
      &&\nonumber \\
   &=&\frac{ \lambda^2}{V}\Theta^2 \int_0^\infty dq\;q^3|A(q)|^2 \left(\frac{k^4 \left(3 q^2-2 k^2\right) \theta (q-k)+q^6 \vartheta
   (k-q)}{8 k^3 q^2}\right) >0 (\mbox{   unless  } |A(q)|=0 ) \nonumber
\end{eqnarray}
This is always positive, therefore $I(k, \Theta)$ is an increasing function of $k$ and it reaches its maximum at $k\rightarrow \infty$. We have 
\begin{equation}
I_{max}(\Theta)=\lim_{k\rightarrow\infty}I(k,  \Theta)=\frac{ \lambda^2}{V}\Theta^2\frac{3}{16\pi} \int_0^\infty dq\;q^3|A(q)|^2
\end{equation}
and the no-pole condition is certainly verified if we impose
\begin{equation}
I_{max}( \Theta)<1.
\end{equation}
Thanks to the latter, we may  estimate  the first Gribov region and exhibit its dependence on $\Theta$. 


\section{Analysis of $\sigma(k,  A, \Theta)$ in four Dimensions}\label{apb}
Let us now address the problem  in $d=4$ dimensions, in the Euclidean setting as previously done. We  work in spherical coordinates $(q, \psi, \alpha, \phi)$ with $q^2= q^\mu q_\mu, \psi,\alpha \in[0,\pi], \phi\in[0, 2\pi]$ and choose $k_\mu=(k,0,0,0)$. Since the computation is Euclidean we are free to choose the reference frame as we like, so we choose the easiest one  and the result will not the depend on the choice. As for the noncommutativity matrix $\Theta$, according to the choice we have made with   \eqn{nctheta}, the only contributing factor will be $\Theta^{12}$, it being
\be
\Theta^{\mu\nu}q_\mu k_\nu=  - \Theta^{12}q_2 k_1= \Theta k q \sin \psi \cos \alpha
\ee
with $\Theta$ a real parameter.  Eq. \eqn{nopole} becomes then
\begin{eqnarray}\label{4dsigm}
\sigma(k, A, \Theta)&=&\frac{4 \lambda^2}{3V}\int_0^\infty \frac{dq}{16\pi^4}\;q^3|A(q)|^2\int_0^{2\pi} d\phi\int_0^\pi \sin{\alpha}\,d\alpha\int_0^\pi \sin^4{\psi}\,d\psi \frac{\sin^2{\left(\frac{1}{2}\Theta q k \sin{\psi}\cos{\alpha} \right)  } }{(k^2+q^2-2k q\cos{\psi})} \nonumber\\
&=&\frac{8\pi \lambda^2}{3V}\int_0^\infty \frac{dq}{16\pi^4}\;q^3|A(q)|^2\int_0^\pi \sin{\alpha}\, d\alpha\int_0^\pi \sin^4{\psi}\, d\psi \frac{\sin^2{\left(\frac{1}{2}\Theta q k \sin{\psi}\cos{\alpha} \right)  } }{(k^2+q^2-2k q\cos{\psi})}\nn
\end{eqnarray}
As for the two-dimensional case,  since   $\frac{\sin^2(x)}{x^2}\leq1$ we 
have
{\small{\begin{eqnarray}
\sigma(k, A, \Theta) &\leq& I(k,  A, \Theta)  =
\frac{8\pi \lambda^2}{3V}\int_0^\infty \frac{dq}{16\pi^4}\;q^3|A(q)|^2\int_0^\pi \sin{\alpha}\,d\alpha\int_0^\pi \sin^4{\psi}d\psi \frac{{\left(\frac{1}{4}\Theta^2 q^2 k^2 \sin^2{\psi}\cos^2{\alpha} \right)  } }{(k^2+q^2-2k q\cos{\psi})} \nonumber\\
&=&\frac{2\pi \lambda^2}{V}\Theta^2 \int_0^\infty \frac{dq}{16\pi^4}\;q^5|A(q)|^2\int_0^\pi \cos^2{\alpha} \sin{\alpha}\,d\alpha\int_0^\pi d\psi \frac{{\left( k^2 \sin^6{\psi}\right)  } }{(k^2+q^2-2k q\cos{\psi})} \nonumber\\
&=&\frac{4\pi \lambda^2\Theta^2}{9V}\int_0^\infty \frac{dq}{16\pi^4}\;q^5|A(q)|^2\int_0^\pi d\psi \frac{{\left( k^2 \sin^6{\psi}\right)  } }{(k^2+q^2-2k q\cos{\psi})} \\
&& \nonumber\\
&=&\frac{ \lambda^2\Theta^2}{36\pi^2 V}\int_0^\infty dq\;q^5|A(q)|^2 \left(
\frac{\left(q^4-5 k^2 q^2+10 k^4\right) \vartheta (k-q)}{16 k^4}+
\frac{k^2 \left(10 q^4-5 k^2 q^2+k^4\right) \vartheta (q-k)}{32 q^6}
\right) \nonumber
\end{eqnarray}
}}
where $\vartheta(x)$ is the Heaviside step function. We now proceed as in 
the previous section, by  analysing the behaviour of the $I(k,A,\Theta)$ as a function of $k$. The derivative of $I(k, A,  \Theta)$ with respect to 
$k$ is given by:
\begin{eqnarray}
\frac{\partial I}{\partial k}&=&\frac{4 \pi^2 \lambda^2\Theta^2}{9V}\int_0^\infty dq\,q^5|A(q)|^2
\left(\frac{10 k^4+q^4 -5 k^2 q^2 }{16 k^4}\delta (k-q)-\frac{10
   k^2 q^4+k^6-5 k^4 q^2}{32 q^6} \delta (q-k) \right. \nonumber\\
   &&\left.+\frac{2 q^8 \left(5 k^2-2 q^2\right) \vartheta
   (k-q)+k^6 \left(-10 k^2 q^2+3 k^4+10 q^4\right) \vartheta (q-k)}{16 k^5 q^6}\right)\nn\\
   &=&\frac{4 \pi^2 \lambda^2\Theta^2}{9V}\int_0^\infty dq\,q^5|A(q)|^2
   \left(\frac{2 q^8 \left(5 k^2-2 q^2\right) \vartheta
   (k-q)+k^6 \left(3 k^4+10 q^4-10 k^2 q^2\right) \vartheta (q-k)}{16 k^5 
q^6}\right)\nonumber \\
   &+&\frac{ \pi \lambda^2\Theta^2}{12V} k^5 \vartheta (k) |A(k)|^2.
\end{eqnarray}
The latter is always strictly positive for $|A(q)|\neq 0$, therefore
 $I(k; A; \Theta)$ is an increasing function of $k$ and its maximum is obtained by performing  the limit $k\rightarrow\infty$:
\begin{equation}
I_{max}( A, \Theta)=\frac{5  \lambda^2\Theta^2}{288 \pi^2 V}\int_0^\infty dq\;q^5|A(q)|^2.
\end{equation}
Therefore the  no-pole condition is attained by imposing
\begin{equation}
I_{max}(A, \Theta)<1
\end{equation}
and the first Gribov region is taken into account in the functional integral  \eqn{fungen} by posing  
\begin{equation}
V(\Omega)=\vartheta(1-I_{max}(A, \Theta)),
\end{equation} 
with $\vartheta$ the Heaviside function. The same conclusions as for the two dimensional case apply: the horizon depends on the square of the non-commutativity parameter and is removed in the commutative limit. The estimate can be probably made more precise by maximising directly the function $\sigma(k; A, \Theta)$, provided one is able to overcome the even more challenging technical difficulties in dealing with the higher-dimensional 
integral 
\eqn{4dsigm}.

\section{ What to expect of an all order computation?}\label{appALLORDERS}
 Here we discuss perspectives on the extension of analysis to include  higher order terms. We start with equation \eqref{eq654}
\begin{eqnarray}
\mathbb{A}^{2}_{kk}&=&\int \frac{d^d q}{(2\pi)^{d}}\, \langle k\vert \mathbb{A}\vert q\rangle\langle q\vert \mathbb{A}\vert k \rangle \nonumber\\
&=&-4\int \frac{d^d q}{(2\pi)^{d}}\sin\left(\frac{1}{2}\Theta_{\rho\sigma}k^{\rho}q^{\sigma}\right)\tilde A_{\mu}(k-q)\frac{q_{\mu}}{q^2}\sin\left(\frac{1}{2}\Theta_{\rho\sigma}q^{\rho}k^{\sigma}\right)\tilde A_{\nu}(q-k)\frac{k_{\nu}}{k^2}.
\end{eqnarray}
One can easily proceed and write a general formal expression for the term of order $n$
\begin{eqnarray}
	\mathbb{A}^{n}_{kk}&=&\int \frac{d^d q_1}{(2\pi)^{d}}\,\cdots  \int \frac{d^d q_{n-1}}{(2\pi)^{d}}\, \langle k\vert \mathbb{A}\vert q_1\rangle\langle q_1\vert \mathbb{A}\vert q_2 \rangle \cdots \langle q_{n-2}\vert \mathbb{A}\vert q_{n-1}\rangle\langle q_{n-1}\vert \mathbb{A}\vert k \rangle \nonumber\\
	&=&(-2i)^n\int \frac{d^d q_1}{(2\pi)^{d}}\,\cdots  \int \frac{d^d q_{n-1}}{(2\pi)^{d}}\,\sin\left(\frac{1}{2}\Theta_{\rho\sigma}k^{\rho}q_1^{\sigma}\right)\sin\left(\frac{1}{2}\Theta_{\rho\sigma}q_1^{\rho}q_2^{\sigma}\right)\cdots \sin\left(\frac{1}{2}\Theta_{\rho\sigma}q_{n-1}^{\rho}k^{\sigma}\right) \nonumber\\
	&&\tilde A_{\mu}(k-q_1)\tilde A_{\sigma_1}(q_1-q_2)\cdots \tilde A_{\sigma_{\nu}}(q_{n-1}-k)
	\frac{q_1^{\mu}q_2^{\sigma_1}\cdots q_{n-1}^{\sigma_{n-1}}}{q_1^2 q_2^2\cdots q_{n-1}^2}\frac{k_{\nu}}{k^2}\nonumber\\
	&=&(-2i)^n \frac{k_{\mu}k_{\nu}}{k^2}\int \frac{d^d q_1}{(2\pi)^{d}}\,\cdots  \int \frac{d^d q_{n-1}}{(2\pi)^{d}}\,\sin\left(\frac{1}{2}\Theta_{\rho\sigma}k^{\rho}q_1^{\sigma}\right)\sin\left(\frac{1}{2}\Theta_{\rho\sigma}q_1^{\rho}q_2^{\sigma}\right)\nonumber\\
	&&\cdots \sin\left(\frac{1}{2}\Theta_{\rho\sigma}q_{n-1}^{\rho}k^{\sigma}\right) \tilde A_{\mu}(k-q_1)\tilde A_{\sigma_1}(q_1-q_2)\cdots \tilde A_{\nu}(q_{n-1}-k)
	\frac{q_2^{\sigma_1}\cdots q_{n-1}^{\sigma_{n-1}}}{q_1^2 q_2^2\cdots q_{n-1}^2}.
\end{eqnarray}
where the Landau gauge property $q_{\mu}A_{\mu}(k-q)=k_{\mu}A_{\mu}(k-q)$  was used. It then follows that the exact ghost form factor is given by
\begin{eqnarray}
	\sigma(k,A) = \sum_n \sigma^{(n)}(k,A),
\end{eqnarray}
where 
\begin{eqnarray}
	\sigma^{(n)}(k,A) &=&\frac{(-2i\lambda)^n}{V} \frac{k_{\mu}k_{\nu}}{k^2}\int \frac{d^d q_1}{(2\pi)^{d}}\,\cdots  \int \frac{d^d q_{n-1}}{(2\pi)^{d}}\,\sin\left(\frac{1}{2}\Theta_{\rho\sigma}k^{\rho}q_1^{\sigma}\right)\sin\left(\frac{1}{2}\Theta_{\rho\sigma}q_1^{\rho}q_2^{\sigma}\right)\nonumber\\
	&&\cdots \sin\left(\frac{1}{2}\Theta_{\rho\sigma}q_{n-1}^{\rho}k^{\sigma}\right) \tilde A_{\mu}(k-q_1)\tilde A_{\sigma_1}(q_1-q_2)\cdots \tilde A_{\nu}(q_{n-1}-k)
	\frac{q_2^{\sigma_1}\cdots q_{n-1}^{\sigma_{n-1}}}{q_1^2 q_2^2\cdots q_{n-1}^2}.\nonumber\\
\end{eqnarray}

Now, we worked semiclassically because we stopped at the lowest nontrivial order, which is the quadratic order $\sigma^{(2)}(k,A)$. But in order to go to higher orders one cannot ignore interactions and the problem must be addressed as discussed around \eqref{1PI}, that is, one must deal with $\sigma(k) = \langle \sigma(k,A) \rangle_{1PI}$. This amounts to compute at each order $n$ the following expression
\begin{eqnarray}
	\sigma^{(n)}(k)&=&\frac{(-2i\lambda)^n}{V} \frac{k_{\mu}k_{\nu}}{k^2}\int \frac{d^d q_1}{(2\pi)^{d}}\,\cdots  \int \frac{d^d q_{n-1}}{(2\pi)^{d}}\,\sin\left(\frac{1}{2}\Theta_{\rho\sigma}k^{\rho}q_1^{\sigma}\right)\sin\left(\frac{1}{2}\Theta_{\rho\sigma}q_1^{\rho}q_2^{\sigma}\right)\nonumber\\
	&&\cdots \sin\left(\frac{1}{2}\Theta_{\rho\sigma}q_{n-1}^{\rho}k^{\sigma}\right) \langle \tilde A_{\mu}(k-q_1)\tilde A_{\sigma_1}(q_1-q_2)\cdots \tilde A_{\nu}(q_{n-1}-k)\rangle_{1PI}
	\frac{q_2^{\sigma_1}\cdots q_{n-1}^{\sigma_{n-1}}}{q_1^2 q_2^2\cdots q_{n-1}^2}.\nonumber\\
\end{eqnarray}
Therefore, one must be able to compute the $n$-point function $\langle \tilde A_{\mu}(k-q_1)\tilde A_{\sigma_1}(q_1-q_2)\cdots \tilde A_{\nu}(q_{n-1}-k)\rangle_{1PI}$, a not so easy task indeed.

The desired results rests on a proper understanding of the behavior of  $\sigma(k)$ as a function of $k$. In the semiclassical quadractic case, analyzed in the paper, the result indicating no qualitative change in the propagator can be traced to the presence of sines in the expression for  $\sigma^{(2)}(k)$, that leads to a stronger bound of the integral suppressing the Gribov effects. The expression for $\sigma^{(n)}(k)$  also displays sines and these should similarly provide stronger bounds in comparison to the non-abelian commutative case. So, one is led to speculate that Gribov effects could be suppressed at higher order as well. But, on the other hand, this will depend on the behavior of the gauge field $n$-point function in a complicated self-consistent way, thus making it difficult to settle for a definite answer as to what happens at higher orders.
\end{appendices}

\end{document}